\documentstyle[aps,eqsecnum]{revtex}
\begin{document}
\draft
\title {Measurement of neutron capture on $^{50}$Ti at thermonuclear
energies}
\author{P.~V.~Sedyshev$^1$, P.~Mohr$^{2,3}$, H.~Beer$^4$,
H.~Oberhummer$^2$, Yu.~P.~Popov$^1$, and W.~Rochow$^5$
}
\address{
        $^1$ Frank Laboratory of Neutron Physics, JINR,
        141980 Dubna, Moscow Region, Russia \\
        $^2$ Institut f\"ur Kernphysik, Technische Universit\"at
        Darmstadt,
        Schlossgartenstr.~9, D-64289 Darmstadt, Germany \\
        $^3$ Institut f\"ur Kernphysik, Technische Universit\"at Wien,
        Wiedner Hauptstr.~8-10, A-1040 Vienna, Austria \\
        $^4$Forschungszentrum Karlsruhe, Institut f\"ur Kernphysik III,
        P.~O.~Box 3640, D-76021 Karlsruhe, Germany \\
        $^5$Physikalisches Institut, Universit\"at T\"ubingen,
        Auf der Morgenstelle 14, D-72076 T\"ubingen, Germany
}
\date{\today}
\maketitle
\begin{abstract}
At the Karlsruhe and T\"ubingen 3.75\,MV Van de Graaff accelerators the
thermonuclear $^{50}$Ti(n,$\gamma$)$^{51}$Ti(5.8\,min) cross section was
measured by the fast cyclic activation technique via the
320.852 and 928.65\,keV $\gamma$-ray lines of the $^{51}$Ti-decay.
Metallic Ti samples of natural isotopic composition and samples of TiO$_2$
enriched in $^{50}$Ti by 67.53\,\%
were irradiated between two gold foils which served as capture standards.
The capture cross-section was measured at the neutron energies 25, 30, 52,
and 145\,keV, respectively. The direct capture cross section was determined
to be 0.387$\pm$0.011 mbarn at 30\,keV. We found evidence for a bound state
s-wave resonance with an estimated radiative width of 0.34\,eV which
destructively interfers with direct capture. The strength of a suggested
s-wave resonance at 146.8\,keV was determined. The present data
served to calculate, in addition to the directly measured Maxwellian
averaged capture cross sections at 25 and 52\,keV, an
improved stellar
$^{50}$Ti(n,$\gamma$)$^{51}$Ti rate in the thermonuclear energy region
from 1 to 250\,keV. The new stellar rate leads at low
temperatures to much higher values than the previously recommended rate,
e.g., at $kT=8$\,keV the increase amounts to about 50\,\%. The
new reaction rate therefore reduces the
abundance of $^{50}$Ti due to s-processing in AGB stars. 

\end{abstract}


% 25.40.Lw: Radiative capture
% 24.50.+g: Direct reactions
% 25.40.Dn: Elastic neutron scattering
%
\section{Introduction}\label{s1}
For a long time it has been known that the solar-system abundances of elements
heavier than iron have been produced by
neutron-capture reactions~\cite{bur57}. However, neutron capture
is also of relevance for abundances of isotopes
lighter than iron especially for neutron-rich
isotopes, even though the bulk of these elements
has been synthesized by charged-particle induced reactions.
The attempts to understand neutron-induced nucleosynthesis require as
important ingredients the knowledge of neutron-capture
rates. The influence of shell effects on neutron capture is
one of the most interesting aspects of neutron capture, especially
since neutron capture in the vicinity of magic numbers is often
a bottleneck in neutron-induced nucleosynthesis. This is the case
also in neutron capture on neutron-rich isotopes
close to the magic proton and neutron numbers $Z=20$ and $N=28$, i.e.,
in the vicinity of the double-magic nucleus $^{48}$Ca.
In particular, the reaction rate of 
neutron capture for Ti-isotopes is of relevance
for isotopic abundance anomalies in silicon carbide (SiC) grains occurring in
carbonecous meteorites~\cite{hop94,gal94}. 
Contrary to most other solar system solids this type of grains has not
been reprocessed an/or homogenized. Therefore, they can be can be
potentially associated with their stellar origin. The main part of 
presolar SiC grains have isotopic compositions implying that they most likely
condensed in the winds of a variety of asymptotic giant
branch (AGB) stars~\cite{gal94}.

The SiC grains show a large spread of $\delta^{50}$Ti, 
the permille deviation from the solar ratio
with $^{48}$Ti as reference isotope, which is evidence for
s-processing~\cite{gal94,lug98}. The very
small neutron capture cross section of the 
neutron magic nucleus $^{50}$Ti makes it
behave as a bottleneck in the s-process path, 
building up a considerable abundance.
The predictions
of AGB models discussed in Refs.~\cite{gal98} show for the
neutron-rich isotopes $^{49,50}$Ti that the deviations from their
solar ratio are essentially in agreement
with their measured values in SiC grains~\cite{lug98}.
However, for $^{50}$Ti the calculated permille deviations from the solar ratio
seem a little too high~\cite{gal94}.
This was one of the motivations to remeasure and reinvestigate the neutron
capture cross section on $^{50}$Ti in the thermonuclear energy range.

In Sect.~II the measurements using the fast cyclic activation technique,
the neutron production and the time-of-flight measurements are described.
The analysis of the thermonuclear capture cross-sections and their
interpretation in form of the non-resonant and resonant capture follows in
Sect.~III. Furthermore, the Maxwellian averaged capture cross section is
derived. Finally, in the last section the experimental results and their
theoretical interpretation are discussed and compared with previous data.
Possible astrophysical consequences of the new neutron capture rate of
$^{50}$Ti with respect to the abundance of this isotope are discussed. 
\section{Measurements} 
The thermonuclear measurements have been carried
out at the Karlsruhe and T\"ubingen 3.75\,MV Van de Graaff accelerators.
In the investigated reaction $^{50}$Ti(n,$\gamma$)$^{51}$Ti(5.8\,min) the
characteristic $\gamma$-ray lines of the $^{51}$Ti product nucleus with
E$_{\gamma}$=320 and 928\,keV served as an excellent signature for the
capture events because of the high accuracy of 0.4\,\% with which the
intensity per decay of the 320\,keV line is known (Table~\ref{tt1}). The
capture cross sections were determined relative to the
$^{197}$Au(n,$\gamma$)$^{198}$Au(2.69517\,d) standard
reaction~\cite{mac82,rat88}, where the 411.8044\,keV $\gamma$-ray line of
the $^{198}$Au decay is known with high precision (Table~\ref{tt1}). 
\subsection{Fast Cyclic Activation Technique}\label{s2} 
The activation
technique \cite{bee80,kae87}, especially the fast cyclic activation has
been described in previous publications \cite{bee94,bee95}. An activation
cycle is characterized by an irradiation and activation counting period.
For the short lived $^{51}$Ti product nucleus it is essential to repeat
these cycles frequently to gain statistics. The time constants for each
cycle are the irradiation time $t_{\rm b}$, the counting time $t_{\rm c}$,
the waiting time $t_{\rm w}$ (the time to switch from the irradiation to
the counting phase) and the total time T=$t_{\rm b}$+$t_{\rm w}$+$t_{\rm
c}$+$t'_{\rm w}$
($t'_{\rm w}$ the time to switch from the counting to the irradiation
phase). 
In the actual $^{50}$Ti measurements the runs were carried out with
$t_{\rm b}$=119.6\,s, $t_{\rm c}$=119.6\,s, the
waiting time $t_{\rm w}$=0.40\,s and the total time T=240\,s.
The decay of the $^{51}$Ti product nuclei during
irradiation and the fluctuations of the neutron beam intensity are 
taken into account by the factor $f_{\rm b}$.  

The accumulated number of counts from a total of $N$ cycles,
$C=\sum_{i=1}^n C_i$, where $C_i$,
the counts after the i-th cycle, are calculated for a chosen irradiation
time, $t_{\rm b}$ is~\cite{bee94}
\begin{equation}
\label{eq1}
C  = \epsilon_{\gamma}K_{\gamma}f_{\gamma}\lambda^{-1}[1-\exp(-\lambda
t_{\rm c})]
\exp(-\lambda t_{\rm w})
 \frac{1-\exp(-\lambda t_{\rm b})}{1-\exp(-\lambda T)} N \sigma_\gamma
{[1-f_{\rm b} \exp(-\lambda T)]}
 \sum_{i=1}^n \Phi_i
 \end{equation}
with
\begin{displaymath}
f_{\rm b}  =  \frac{\sum_{i=1}^n \Phi_i \exp[-(n-i)\lambda T]}{
\sum_{i=1}^n \Phi_i} \quad .
\end{displaymath}
The following additional quantities have been defined;
$\epsilon_\gamma$: Ge-efficiency, $K_\gamma$:
$\gamma$-ray absorption, $f_\gamma$: $\gamma$-ray intensity per decay,
$N$: the thickness (atoms per barn) of target
nuclei, $\sigma_\gamma$: the capture cross section, $\Phi_i$: the neutron flux
in the i-th cycle. The
quantity $f_{\rm b}$ is calculated from the registered flux history of a $^6$Li
glass monitor.

The activity of $^{198}$Au is additionally counted after
the end of the cyclic activation consisting of $n$ cycles using
\begin{eqnarray}
\label{eq2}
C_n=\epsilon_\gamma K_\gamma f_\gamma \lambda^{-1} [1-\exp(-\lambda T_{\rm M})]
\exp(-\lambda T_{\rm W})
[1-\exp(-\lambda t_{\rm b})] N \sigma_\gamma f_{\rm b} \sum_{i=1}^n \Phi_i \quad.
\end{eqnarray}
Here $T_{\rm M}$ is the measuring time of the Ge-detector and $T_{\rm W}$
the time
elapsed between the
end of cyclic activation and begin of the new data acquisition.
\subsection{Neutron production, time-of-flight measurements}
Neutrons were generated by the $^7$Li(p,n) and T(p,n) reactions.
Measurements with neutron spectra of mean neutron energies of 25, 30, 52,
and 145\,keV were carried out by adjusting with the accelerator the
appropriate proton energy. The energy spread of the neutron spectra
produced is dependent on the energy loss of the protons in the target
layer. For proton energies close to reaction threshold the target layers
were chosen thick enough that the energy loss of the protons reached
the reaction threshold within the layer. This makes the neutron spectrum
independent of the target thickness. Using the $^7$Li(p,n) and T(p,n)
reactions in this way we generated quasi-Maxwellian spectra with
thermal energies kT=25 and 52\,keV, respectively~\cite{bee80,kae87,rat88}.
The spectrum with a mean neutron energy of 30~keV was obtained via the
$^7$Li(p,n) reaction with a proton energy of 1882\,keV very close to the
reaction threshold energy E$_{th}$=1881\,keV. With this spectrum it was
possible to perform a measurement in between the resonances, where the
16.99\,keV p-wave resonance~\cite{mug81} is most critical.
A typical TOF spectrum at $0^{\circ}$ and the
respective neutron distribution are shown in Fig.~\ref{f2}.
With a proton energy much higher than the reaction threshold energy of
the $^7$Li(p,n) reaction and a thin Li-target the neutron energy
distribution  is dependent on the thickness of the Li layer. The spectrum
at 145\,keV (Fig.~\ref{f22}) was generated using Li-targets of
2.5$\,\mu$m.

The Li-targets were carefully prepared. Metallic Li was evaporated onto
1.5\,mm thick copper backing. The thickness of the Li layer is determined
during evaporation. For each new target also a new copper backing is used
because of Li diffusion into the copper. Targets of good quality were always
freshly made before use because long term storage under liquid nitrogen
already leads to slight damage of the metallic Li through the onset of
oxidation.

The proton energy and the thickness of the Li target were checked by a
a time-of-flight measurement before an activation run was started.
In the flight time
measurements typically a pulse width of 10 ns and a repetition rate of 1
MHz was used. The neutrons were counted by a $^6$Li glass detector (2.54 cm
dia $\times$ 1\,mm) at a flight path of 91\,cm.

To determine the neutron distribution at the sample position
time-of-flight spectra normalized to the proton beam current were
recorded at zero degree and at different angles in steps of 5
degrees. These spectra are needed for calculating the angle integrated
spectrum at the sample position by Montecarlo methods, where the extended
Li-target of 6\,mm diameter and the mean free path of the neutrons through
the sample are taken into account. The neutron spectra shown in the
figures have been determined in this way.

During the activation runs beam currents up to 100 $\mu$A could be applied
without damage of the Li-layer because of an effective water cooling
system. For the kT=52\,keV measurement with the tritium target a beam
current of 12\,$\mu$A was used, and no significant deterioration of the
tritium target was observed during the experiment.
In the activations four spectra were continuously recorded, an 8K
spectrum of the relative neutron flux $vs$ time from the neutron monitor,
an 8K spectrum of the proton beam current $vs$ time, an 8K $\gamma$-ray
spectrum from the Ge-detector and an 1K pulse height spectrum of the
$^6$Li glass monitor. For safety every two hours a backup was stored
from these spectra automatically. From the ratio neutron flux and proton
beam the actual quality of the target was controlled and from the $^6$Li
glass pulse height spectrum the threshold for $\gamma$-ray background
suppression was checked.

\section{Analysis}
\subsection{Thermonuclear capture cross-sections}
Eqs.~(\ref{eq1}) and (\ref{eq2}), respectively, contain the
quantities $\sigma_\gamma$
and the total neutron flux $\sum_{i=1}^n \Phi_i$. The
unknown capture cross-section of $^{50}$Ti is measured relative
to the well-known standard cross section of $^{197}$Au~\cite{mac82,rat88}.
As the metallic natural Ti and $^{50}$TiO$_2$ samples, respectively, to be
investigated are characterized by a finite thickness it is necessary to
sandwich the sample by two comparatively thin gold foils for the
determination of the effective neutron flux at sample position. The
activities of these gold foils were counted also individually after
termination of the cyclic activation. The effective count rate of gold was
obtained from these individual rates as well as from the accumulated gold
count rate during the cyclic activation run.
Therefore, the effective neutron flux at sample position was determined
in two ways by the gold activation according to the Eqs.~(\ref{eq1})
and (\ref{eq2}). Using Eq.~(\ref{eq1}) has the advantage that saturation
effects in the gold activity for irradiations over several days are
avoided~\cite{bee94}.

The efficiency determination of the 60\,\% HPGe-detector has been performed
with calibrated
radioactive sources and the detector simulation program GEANT~\cite{geant}
as already reported elsewhere~\cite{mohr98}. The $\gamma$-ray
absorption in the samples was calculated using tables published by Storm and
Israel~\cite{sto70} and Veigele~\cite{vei73}.
The half-lives and the $\gamma$-ray intensities per decay of $^{51}$Ti
and $^{198}$Au given in Table~\ref{tt1} were taken from \cite{fir96}.

Table~\ref{tt2} gives a survey of the sample weights and the measured
$^{50}$Ti capture cross-sections. The metallic Ti and the $^{50}$TiO$_2$
samples of 6\,mm diameter were sandwiched by thin gold foils of the same
dimensions. At 25\,keV neutron energy measurements were carried out with
titanium oxid sample masses between
40 and 100\,mg.  
The samples of the $^{50}$TiO$_2$ powder were heated to
1000$^{\rm o}$\,C for 3 hours to get stable self-supporting
tablets. No measurable weight loss was observed. 
Only thin samples were used to minimize corrections for neutron scattering
or self-absorption effects, e.g. from oxygen, the Ti isotopes and from the
gold foils back into the sample. Estimations show that these effects are
below 2\% \cite{bee80}. 
Three runs were carried
out with metallic Ti foils of natural isotopic compositon
(Table~\ref{tt2}). The largest multiple scattering effects would have
been expected from oygen. However the runs with the metallic Ti samples
and the runs with the $^{50}$TiO$_2$ samples show no measurable
differences.
In Fig.~\ref{f3} the accumulated
$\gamma$-ray intensity from one of the $^{50}$Ti activations is shown. The
relevant $\gamma$-lines are well isolated on a low level of background counts.

The following systematic uncertainties were combined by quadratic error
propagation; Au standard cross section: 1.5-3\,\%, Ge-detector efficiency:
2\,\%, $\gamma$-ray intensity per decay: 0.5\,\% for the $^{51}$Ti and
0.1\,\%
for the $^{198}$Au decay, divergence of neutron beam: 2-7\,\%, factor
$f_{\rm b}$: 1.5\,\%, sample weight: $<$0.5\,\%.
\subsection{Direct capture}
Our measurement in between the region of resonances, at 20 to 40\,keV is
interpreted as direct capture (DC). The size of this cross section shows
that
it is not a negligible effect. Because of the selectivity of our
activation method only capture in $^{50}$Ti can contribute to this
value. The contributions from the tails of
$^{50}$Ti resonances are estimated to be not more than 0.005 mbarn and are
neglected. This estimate is in agreement with the 
value which can be derived from the respective value for thermal capture
given in Ref.~\cite{mug81}.  Theoretical
estimations suggest that we have to treat our
measured direct capture value as pure s-wave capture with a 1/v energy
dependence. A p-wave contribution would be, as a rough calculation shows,
at 30\,keV $\le$ 17.5\,$\mu$barn. If we assume that the thermal capture 
is also essentially direct capture we can make a comparison. The 1/v
extrapolation of our measured
direct capture cross section to thermal energy gives 0.421$\pm$0.012 barn
which is, however, significantly larger than the measured value of
0.179$\pm$0.003
barn~\cite{mug81}. But if the bound s-wave resonance at -16.8\,keV
specified in the compilation of Mughabghab et al.~\cite{mug81}
exists destructive interference of resonance capture could occur.
As the first positive resonance lies at 56.5\,keV and the s-wave
level spacing is estimated to be D$_0$=125$\pm$70\,keV~\cite{mug81},
this bound state resonance seems to be not unlikely. A capture
cross section of this resonance at 25.3\,meV of 0.051\,barn
which corresponds to a radiative width of this resonance of
0.34\,eV can reduce the direct capture to the observed thermal value.

A theoretical DC calculation was performed using a
realistic neutron-nucleus folding potential \cite{ca48,ti44}
and spectroscopic factors of $^{51}$Ti from the (d,p) experiment
of Ref.~\cite{kocher72}. The result of the DC calculation
overestimates our experimental value at 30\,keV by a factor of 2;
the calculated cross section depends on the choice of the potential
parameters and the uncertainties of the spectroscopic factors. The
resulting
uncertainty of the calculated value
is also roughly a factor of 2.

A further indication of the strong influence of the subthreshold state
on the thermal capture cross section comes from the comparison of the
experimental \cite{ndsti50} and theoretical branching ratio. In the DC
calculations the strongest transition is the ground state
($3/2^-$) transition
with a branching of about 60\% (independent on the choice of the
potential parameters). However, experimentally the transition to the
first excited $1/2^-$ state at 1167\,keV is dominating with 35\%,
and the ground state branching is almost negligible (4.2\%).
It should be noted that this experimental branching ratio is very
surprising because in the neighboring nucleus $^{48}$Ca the experimental
and the theoretical branching ratios agree well and show a dominating
transition to the first $3/2^-$ state. The disappearance of the
ground state transition in $^{50}$Ti can be explained by
destructive interference with the proposed $1/2^+$ subthreshold
state at -16.8\,keV. Furthermore, a strong branching ratio of this
subthreshold state can be predicted from the interference effect
which mainly influences the ground state transition at thermal energies.

\subsection{Resonance capture at 145\,keV}
The $^{50}$Ti capture cross section has been previously measured by Allen
et al.~\cite{all77} using the time-of-flight technique. These data have
been essentially adopted in the compilation of Mughabghab et
al.~\cite{mug81} except for the g$\Gamma_n\Gamma_\gamma/\Gamma$ value
of the 97.6\,keV p-wave resonance which was changed
from 0.77 to 0.16\,eV. In an additional
publication Allen et al.~\cite{all79} corrected the $\Gamma_\gamma$ value
of the 56.5\,keV s-wave resonance from 1.1$\pm$0.2\,eV to 0.2$\pm$0.2\,eV and
estimated also the $\Gamma_\gamma$ widths for the s-wave resonances
at 146.8 and 184.9\,keV to be $<$ 0.1 and 0.9$\pm$0.3\,eV, respectively.
These revisions were necessary as the
Oak Ridge capture detection setup had turned out to be very sensitive to
scattered neutrons resulting in a high prompt (time dependent) capture
$\gamma$-ray background which requires large corrections to the
capture areas of the s-wave resonances with a large ratio of neutron
scattering and capture width~\cite{alle79}. The areas of the s-wave
resonances at 56.5 and 184.9\,keV had to be corrected for prompt
background contributions of 87 and 73\,\%~\cite{all79}, respectively.
For the energy interval 100 to 150\,keV Allen et al.~\cite{all77,all79}
report an average capture cross section of 0.54$\pm$0.08\,mbarn which could
contain contributions of the 146.8\,keV s-wave resonance. However, Allen's
average cross section is, it seems to be, only derived from the two
identified resonances in this energy region. The resonance strenghts
of the 101.4 and 120.6\,keV resonances exactly add up to the given
average cross section. In an effort to obtain
an estimate for the strength of the 146.8 keV resonance we have
carried out a measurement with a neutron spectrum of 145$\pm$20\,keV.
Our measured average cross section at 145 keV of 0.650$\pm$0.029\,mbarn
consists
of a direct contribution of 0.175$\pm$0.005\,mbarn extrapolated from the
measured value at 30\,keV, and the resonance contributions from the known
101.4, 120.6, and 185.6\,keV resonances and the 146.8\,keV s-wave
resonance
to be investigated. The average capture cross section of a narrow resonance
compared to the neutron spectrum used can be described in the following
way
\begin{eqnarray}
\label{eq3}
\bar{\sigma}_{\rm res}=\frac{\int \sigma_{\rm BW}\Phi(E)dE}{\int \Phi(E)dE}
= A_{\gamma} \frac{\Phi(E_{\rm res})}{\int \Phi(E)dE}
=A_{\gamma} \Phi_{\rm norm}(E_{\rm res})
\end{eqnarray}
where $\sigma_{BW}$ is the Breit-Wigner formula for the resonace capture,
$\Phi(E)$ the experimental neutron spectrum and A$_{\gamma}$=
(2$\pi^2$/k$^2$) g$\Gamma_{\gamma} \Gamma_n/\Gamma$ the resonance area
with k the wave number, g=(2J+1)/[2(2I+1)] the statistical spin factor with
J the compound and I the nuclear spin, and $\Gamma_{\gamma}$,
$\Gamma_n$ and $\Gamma$ the radiation, neutron and total widths,
respectively.
After subtraction of the direct capture and the resonance capture
from the 101.4, 120.6, and 185.6\,keV resonances, where the resonance
capture contribution of the 120.6\,keV resonance of 0.15$\pm$0.07 mb is
comparable in magnitude with the direct capture part, we obtain
for the average resonance capture of the 146.8\,keV s-wave
resonance 0.22$\pm$0.13\,mb which corresponds with a radiation width
g$\Gamma_{\gamma} \Gamma_n/\Gamma$=0.37$\pm$0.24\,eV. The dominant
uncertainties in this analysis come from the neutron
spectrum determinations
$\Phi_{\rm norm}(E_{\rm res}=120.6$\,keV)=
(7$\pm$3)$\times$10$^{-3}$\,keV$^{-1}$,
$\Phi_{\rm norm}(E_{\rm res}=146.8$\,keV)=
(21$\pm$5)$\times$10$^{-3}$\,keV$^{-1}$, and
$\Phi_{\rm norm}(E_{\rm res}=185.6$\,keV)=
(4$\pm$5)$\times$10$^{-3}$\,keV$^{-1}$.
The uncertainties of the capture
areas for the 101.4, 120.6 keV and 185.6\,keV resonances
have been taken from Allen et al.~\cite{all77,all79}.
\subsection{Maxwellian averaged capture cross
sections and reaction rate factors}
\label{s3}
The Maxwellian averaged capture cross section (4.0$\pm$0.5\,mbarn at
kT=30\,keV) recommended until now in literature~\cite{bk87,bvw92} has been
calculated using the resonance parameters given in the tables of
Mughabghab et al.~\cite{mug81} which are based on the reported
resonance strengths in Ref.~\cite{all77}. Direct capture was neglected
and the revised radiation widths of the s-wave resonances \cite{all79}
were not considered. In order to compare our measured Maxwellian
averaged capture (MAC) cross sections
at kT=25\,keV and 52\,keV, respectively with the corresponding values
calculated from the resonance parameters we have to take into account
the above mentioned changes. In Table \ref{tt3} we calculated first
the MAC cross sections at kT=25\,keV and 52\,keV
using the resonance parameters from Ref.~\cite{mug81} (CALC1),
then we took into account the revised s-wave resonance
widths reported in Ref.~\cite{all79}(CALC2), we included in CALC3
the direct capture contribution and in CALC4 our determined radiation
width of the 146.8\,keV resonance. The direct capture contribution
(CALC3) was calculated using our measured cross section value at 30\,keV 
and the known 1/v energy dependence, where the MAC cross section
$<\sigma$ v$>$/v$_T$(kT=E$_n$)=$\sigma^{DC}_{\gamma}$(E$_n$).
The biggest changes are
caused by considerably smaller radiation widths for the s-wave resonances,
especially for the 56.5\,keV resonance (CALC2) and the inclusion of the
direct s-wave capture contribution (CALC3). Finally, CALC4 is distincly
in better agreement with our experimental MAC cross sections than CALC1.
In Fig.~\ref{f4} the results of CALC1 and CALC4 $vs$ kT are plotted
together
with our measurements at kT=25 and 52\,keV and the 30\,keV MAC cross section
reported in Refs.~\cite{mug81,all77}. Below 17\,keV where there are no
$^{50}$Ti resonances the direct capture starts to provide the dominant
contribution. This part of the capture cross section is well determined by
our direct capture measurement at 30\,keV and the known temeprature
dependence. Theoretical determinations
are also shown. The curves WHFZ and SMOKER refer to these Hauser-Feshbach
calculations given in \cite{woo78,CTT91}. In the SMOKER calculation
\cite{CTT91} a pure 1/v temperature dependence was assumed, and the
temperature dependence of the WHFZ calculation \cite{woo78} is practically 
also 1/v. As there are no $^{50}$Ti resonances below neutron
energies of 17\,keV these
calculations overestimate the MAC cross section in the temperature
range below kT=20\,keV.     

The final MAC cross sections and reaction rate factors (CALC4) are listed
{\rm vs} temperature kT in Table \ref{tt4}.
\subsection{Conclusions}
In our experiment we found the direct s-wave capture component
characterising this magic shell isotope. Our measured direct capture
cross section at 30\,keV is one of the few examples, where the
measurement of the direct part in the presence of strong resonance capture
was successful. The measured thermal cross section
is most likely the product of destructive interference with a bound s-wave
resonance the existence of which was already suggested \cite{mug81}. The
present investigations yields confirmative arguments. The estimated
resonance strength of 0.34\,eV is in accord with the radiation widths of the
positive s-wave resonances determined by Allen et al.~\cite{all79} and in
this work. The inclusion of a direct capture component
into the calculation of the stellar reaction rate of $^{50}$Ti is of
importance. It compensates the decrease of the MAC cross section at
kT=25\,keV due to the necessary reduction of the s-wave radiation width
especially of the 56.5\,keV resonance so that the calculation from the
resonance parameters remains in agreement with our direct MAC cross
section measurement. This may be considered also as an evidence for the
consistence of the present data.
It is also important to note that the stellar $^{50}$Ti reaction rate 
is not constant because the capture cross section has a
non 1/v dependence in the thermonuclear temperature range important for the
investigation of the isotopic titanium anomalies which are supposed to be
of s-process origin. For a further
improvement of the $^{50}$Ti MAC cross section a new time-of-flight
measurement is recommended in order to determine more accurately especially
the s-wave radiation widths.

The direct capture contribution becomes a significant part of the MAC cross
section at low thermonuclear energies, i.e.~at kT= 8\,keV, where the
$^{13}$C($\alpha$,n) neutron source is ignited
and s-processing takes place in low-mass AGB stars~\cite{gal98}. Compared
to the old data (CALC1) the direct capture increases the MAC cross section
at kT= 8\,keV by about 50\,\% (see Fig.~\ref{f4}).
This also reduces the calculated abundance of $^{50}$Ti in AGB
models~\cite{gal94,gal98}. As was stated in Ref.~\cite{gal94} such a
reduction could give better agreement with the measured values of the solar
ratio with $^{50}$Ti as reference isotope in SiC grains~\cite{hop94}.

\section*{Acknowledgements}
The interest and support of Prof. G. J. Wagner of the Universit\"at T\"ubingen
is gratefully acknowledged. We would like to thank the technicians of the
Van de Graaff accelerators
M. Brandt and G. Rupp, respectively, and the Karlsruhe Van de Graaff staff
members E.P. Knaetsch, D. Roller, and W. Seith for their help and support
of the experiment especially in the preparation of the metallic
Li-targets.
We thank the  Fonds zur F\"orderung der wissenschaftlichen Forschung in
\"Osterreich (project S7307-AST), the Deutsche Forschungsgemeinschaft
(DFG) (project Mo739/1-1), and Volkswagen-Stiftung (Az: I/72286)
for their support.

\clearpage
\begin{table}
\caption{\label{tt1} Sample characteristics and decay properties of the
product nuclei $^{51}$Ti and $^{198}$Au}
\begin{center}
\begin{tabular}{ccccccc}
Isotope&Chemical &Isotopic composition & Reaction & $T_{1/2}$ & $E_\gamma$ &
Intensity per decay\\
  &form   &  (\%)   &   &   &  (keV)    & (\%) \\
\hline
$^{50}$Ti&TiO$_2$& 2.87(46), 2.54(47), 23.97(48), &
$^{50}$Ti(n,$\gamma$)$^{51}$Ti&5.76\,min& 320.0852
&93.1$\pm$0.4$^a$\\
&& 3.09(49), 67.53(50)   &&&928.65 &6.90$\pm$0.37$^a$\\
&metallic&   natural             &&&       &             \\
&&                       &&&       &             \\
$^{197}$Au&metallic& 100&$^{197}$Au(n,$\gamma$)$^{198}$Au&2.69\,d&
411.8047&95.50$\pm$0.096\\
\end{tabular}
\end{center}
{$^a$ Reference~\cite{fir96}}
\end{table}

\begin{table}
\unitlength1cm
\caption{\label{tt2} Sample weights and experimental $^{50}$Ti
capture cross-sections at thermonuclear energies}
\begin{center}
\begin{tabular}{ddddddddd}
Mean neutron  & Mass of Au &  Mass of&Chemical  & Mass of Au &
Isotopic&
$\sigma_\gamma$ &\multicolumn{2}{c}{Uncertainty}\\
energy& back side& sample&form& front side&composition & (mbarn) &
statistical&total\\
(keV)   &   (mg)  &  (mg) & &(mg)  &of Ti sample&  &(\%)&(\%)\\
\hline
kT=25 &  16.253&95.447&TiO$_2$& 16.253& enriched & 4.117 &0.50&3.98\\
      &  16.640&49.185&TiO$_2$& 16.580& enriched & 3.134 &0.32&6.43\\
      &  16.070&40.900&Ti & 16.110& natural  & 3.550 &0.50&2.84\\
      &  16.033&40.892&Ti & 16.063& natural  & 3.643 &0.72&2.84\\
      &  15.857&49.185&TiO$_2$& 15.843& enriched & 3.468 &0.72&5.13\\
      &  16.600&95.447&TiO$_2$ & 16.593& enriched & 3.267 &0.50&4.10\\
      &  16.433&40.830&Ti & 16.413& natural  & 3.795 &0.98&3.18\\
\cline{6-8}
\multicolumn{6}{r}{Average} &                   3.612$\pm$0.095&\\
\hline
kT=52 &  16.386&95.447&TiO$_2$ & 16.590& enriched & 2.536 &0.72&4.41\\
      &  16.587&49.185&TiO$_2$ & 16.567& enriched & 2.857 &1.21&4.19\\
\cline{6-8}
\multicolumn{6}{r}{Average} &                   2.736$\pm$0.113&\\
\hline
$30 \pm 5$ &16.420&95.447&TiO$_2$ & 16.405& enriched & 0.390 &1.04&2.80\\
           &16.067&49.185&TiO$_2$ & 16.053& enriched & 0.383 &3.03&4.65\\
           &16.010&49.185&TiO$_2$ & 16.017& enriched & 0.421 &1.03&4.13\\
           &16.507&95.447&TiO$_2$ & 16.500& enriched & 0.181 &11.91&12.86\\
           &16.386&95.447&TiO$_2$ & 16.323& enriched & 0.450 &5.62&6.21\\
\cline{6-8}
\multicolumn{6}{r}{Average} &                   0.387$\pm$0.011&\\
\hline
$145 \pm 16$&16.173&49.185&TiO$_2$ & 16.200& enriched & 0.605&2.94&7.57\\
            &16.320&95.447&TiO$_2$& 16.327& enriched & 0.679&0.71&4.70\\
            &16.127&49.185&TiO$_2$& 16.110& enriched & 0.586&1.53&6.60\\
\cline{6-8}
\multicolumn{6}{r}{Average} &                   0.650$\pm$0.029&\\
\end{tabular}
\end{center}
\end{table}

\begin{table}
\caption{\label{tt3} Comparison of the measured Maxwellian averaged
capture (MAC)
cross sections with calculations from the resonance parameters.}
\begin{center}
\begin{tabular}{cccccc}
kT&\multicolumn{5}{c}{$<\sigma$ v$>$/v$_T$ (mbarn)} \\
(keV)&&&&&\\
 & CALC1$^a$ & CALC2$^b$ & CALC3$^c$&CALC4$^d$ & Present measurement  \\
\hline
8  &1.54$\pm$0.19& 1.47$\pm$0.18&2.23$\pm$0.19&2.23$\pm$0.19&\\
25 &3.82$\pm$0.48&
3.02$\pm$0.38&3.45$\pm$0.38&3.45$\pm$0.38&3.612$\pm$0.095\\
30 &4.00$\pm$0.50$^e$&3.18$\pm$0.40&3.57$\pm$0.40&3.58$\pm$0.40&\\
52 &3.51$\pm$0.44&
2.84$\pm$0.36&3.14$\pm$0.36&3.17$\pm$0.36&2.736$\pm$0.113\\
\end{tabular}
\end{center}
{$^a$ resonances taken from Reference~\cite{mug81}}\\
{$^b$ same as in CALC1, but revised
parameters of the s-wave resonances at 56.5, 146.8,
and 184.9 keV from Reference~\cite{all79}}\\
{$^c$ same as in CALC2, but a direct s-wave capture
contribution measured in this work was included}\\
{$^d$ same as in CALC3, but with the resonance strength of
the 146.8\,keV resonance determined in this work} \\
{$^e$ this value is reported in Refs.~\cite{mug81,all77} and was
adopted in the compilations Refs.~\cite{bk87,bvw92}}
\end{table}

\begin{table}
\caption{\label{tt4} The Maxwellian averaged
capture (MAC) and stellar reaction rate
from CALC4 between 1 and 250\,keV.}
%\begin{center}
\begin{tabular}{ccc}
kT&$<\sigma$ v$>$/v$_T$ &Stellar rate factor\\
(keV)&(mbarn)&(10$^4$cm$^3$mol$^{-1}$s$^{-1}$)\\
\hline
1  &2.13$\pm$0.06&5.66\\
2  &1.55$\pm$0.04&5.84\\
3  &1.54$\pm$0.05&7.09\\
4  &1.76$\pm$0.09&9.34\\
5  &1.97$\pm$0.13&11.71\\
6  &2.11$\pm$0.16&13.74\\
7  &2.18$\pm$0.17&15.38\\
8  &2.23$\pm$0.19&16.75\\
10 &2.30$\pm$0.20&19.31\\
12 & 2.41$\pm$0.23&22.24\\
15 & 2.68$\pm$0.27&27.60\\
18 & 2.97$\pm$0.31&33.53\\
20 & 3.15$\pm$0.33&37.42\\
22 & 3.29$\pm$0.35&41.08\\
25 & 3.45$\pm$0.38&45.92\\
27 & 3.52$\pm$0.39&48.69\\
30 & 3.58$\pm$0.40&52.14\\
35 & 3.58$\pm$0.40&56.27\\
40 & 3.49$\pm$0.39&58.78\\
45 & 3.37$\pm$0.38&60.15\\
50 & 3.23$\pm$0.37&60.72\\
52 & 3.17$\pm$0.36&60.79\\
55 & 3.08$\pm$0.35&60.75\\
60 & 2.93$\pm$0.33&60.40\\
65 & 2.79$\pm$0.32&59.81\\
79 & 2.65$\pm$0.30&59.05\\
80 & 2.41$\pm$0.27&57.23\\
90 & 2.19$\pm$0.25&55.21\\
100& 2.00$\pm$0.22&53.15\\
110& 1.83$\pm$0.20&51.10\\
120& 1.68$\pm$0.19&49.10\\
150& 1.34$\pm$0.15&43.57\\
180& 1.09$\pm$0.12&38.81\\
200& 0.958$\pm$0.103&36.03\\
250& 0.719$\pm$0.077&30.25\\
\end{tabular}
%\end{center}
\end{table}

\newpage

\begin{figure}[t]
% m31_n30.eps
\caption{\label{f2} Left: Time-of-flight (TOF) neutron spectrum
from the $^7$Li(p,n) reaction 1\,keV above reaction threshold.
The position of the $^{50}$Ti p-wave resonance at 16.99\,keV is shown
together with a
shaded area which is the uncertainty of the TOF measurement.
Right: Angle integrated neutron spectrum
from the $^7$Li(p,n) reaction 1\,keV above reaction threshold.
}
\end{figure}

\begin{figure}[t]
% spec160.eps
\caption{\label{f22} Neutron spectrum with a mean energy of 145.1\,keV.
The distribution has been generated from time-of-flight
spectra measured at different angles. The integration has been performed
by Monte Carlo methods.}
\end{figure}

\begin{figure}[t]
% ti50a.eps
\caption{\label{f3} Accumulated intensities of the
$^{51}$Ti and $^{198}$Au $\gamma$-ray decay lines
from the activation with a 95.447\,mg TiO$_2$ sample sandwiched
by two Au foils using a neutron spectrum
with a mean energy of kT=25\,keV.}
\end{figure}

\begin{figure}[t]
% calc.eps
\caption{\label{f4} Comparison of our measured MAC cross sections (full
black circles) with the calculations CALC1 (dashed line) and CALC4 (solid
line) from the resonance parameters. The 30\,keV MAC cross section as
reported in literature is shown as well. For the calculation CALC4 also
the uncertainty is given. The influence of the direct capture contribution 
is clearly seen below 10\,keV.
Additionally theoretical determinations are given.
The labels WHFZ~\protect\cite{woo78} (dash-dot-dotted curve) and
SMOKER~\protect\cite{CTT91} (dash-dotted curve)
refer to these Hauser-Feshbach calculations. 
} 
\end{figure}

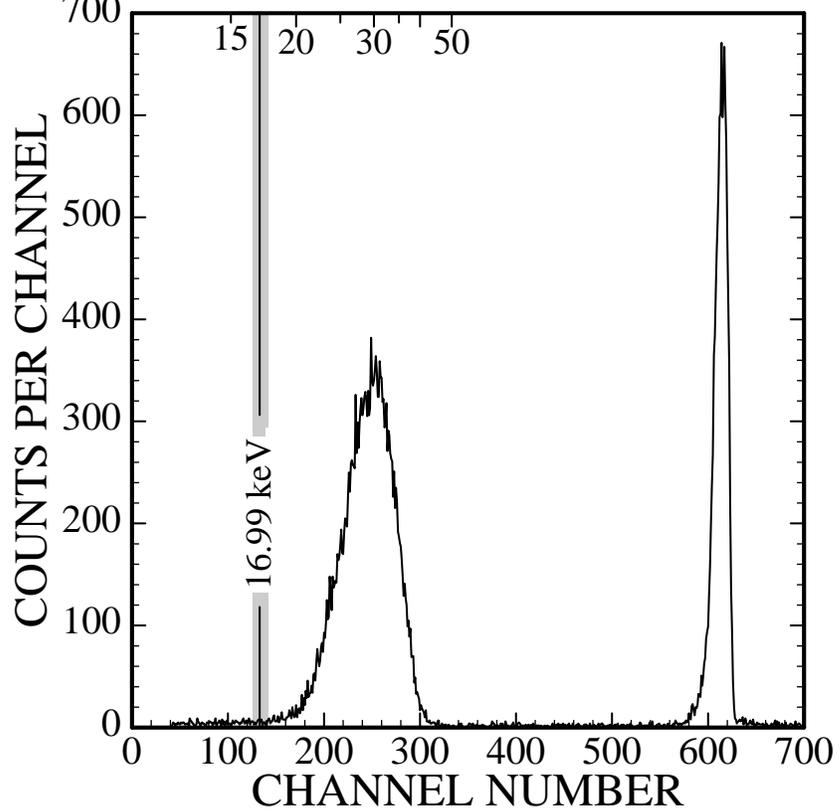
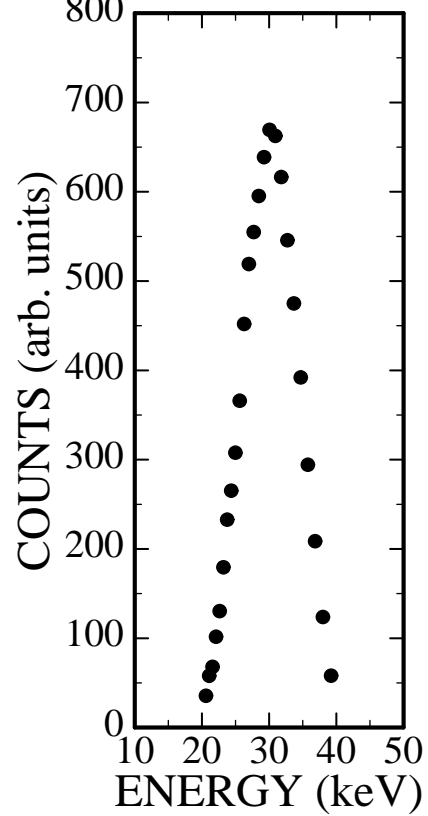
\end{document}